\documentclass{article}
\usepackage{setspace}

\usepackage{PRIMEarxiv}

\usepackage[utf8]{inputenc} 
\usepackage[T1]{fontenc}    
\usepackage{url}            
\usepackage{booktabs}       
\usepackage{amsfonts}       
\usepackage{nicefrac}       
\usepackage{microtype}      
\usepackage{lipsum}
\usepackage{fancyhdr}       
\usepackage{graphicx}       
\usepackage{bm}
\usepackage{multirow}
\usepackage{adjustbox}
\usepackage{float}
\usepackage{parskip}
\usepackage{algorithm,algpseudocode}

\graphicspath{{media/}}     
\RequirePackage{amsthm,amsmath,amsfonts,amssymb}
\RequirePackage[round,authoryear]{natbib}
\RequirePackage[colorlinks,citecolor=blue,urlcolor=blue,linkcolor=blue]{hyperref}

\title{Using Echo State Networks to Inform Physical Models for Fire Front Propagation
}

\author{
  Myungsoo Yoo\thanks{Author of correspondence} \\
  University of Missouri \\
  \texttt{mym4v@mail.missouri.edu} \\
   \And
  Christopher K. Wikle \\
  University of Missouri \\
  \texttt{wiklec@missouri.edu} \\
}

\begin{document}
\raggedbottom
\maketitle
\begin{abstract}
Wildfires can be devastating, causing significant damage to property, ecosystem disruption, and loss of life.   Forecasting the evolution of wildfire boundaries is essential to real-time wildfire management. To this end, substantial attention in the wildifre literature has focused on the level set method, which effectively represents complicated boundaries and their change over time. Nevertheless, most of these approaches rely on a heavily-parameterized formulas for spread and fail to account for the uncertainty in the forecast. The rapid evolution of large wildfires and inhomogeneous environmental conditions across the domain of interest (e.g., varying land cover, fire-induced winds) give rise to a need for a model that enables efficient data-driven learning of fire spread and allows uncertainty quantification.  Here, we present a novel hybrid model that nests an echo state network to learn nonlinear spatio-temporal evolving velocities (speed in the normal direction) within a physically-based level set model framework. This model is computationally efficient and includes calibrated uncertainty quantification.  We show the forecasting performance of our model with simulations and two real data sets - the Haybress and Thomas megafires that started in California (USA) in 2017.
\end{abstract}


\section{Introduction}\label{sec:1}
Wildfires can negatively affect natural and human ecosystems through loss of life, property damage, emission of air pollutants and impacts on water quality, among others \citep[e.g.,][]{brotons, Hohner}. Indeed, wildfires can continue to pose a risk even after they are extinguished. For instance, \citet{cannon} raised public awareness of the impact of the post-fire debris flow. As stake holders consider tradeoffs between managing ecosystems and preventing loss of life and property, having as much information to plan wildfire suppression tactics as early as possible is important. This requires an efficient and reliable forecasting model for the evolution of wildfires.

The level set method proposed by \citet{osher2} has become one of the most useful approaches for modeling the evolution of wildfire fronts in operational fire models \citep[e.g.,][]{miller2015spark}. \citet{osher} summarized various implementations of the level set method, and \citet{gibou} provided a recent review. In the application to wildfire, the level set method frequently relies on heavily-parameterized quasi-empirical formulations of the wildfire front velocity, such as the Rothermel model \citep{rothermel}. Such models can be suitable and perform well under homogeneous environmental and atmospheric conditions. For example, \citet{dabrowski2022towards} used the Rothermel model with an unknown parameter to inform velocity in the level set method by using an ensemble Kalman filter (EnKF).  Similarly, \citet{mallet} and \citet{Alessandri} simplified the velocity through formulas proposed by \citet{fendell} and \citet{lo}, respectively. 
None of these approaches utilizes a data-driven approach to learn the spatially and temporally varying front velocity over the entire domain, which is required for realistic front propagation in complex environments, especially where large fires create their own ``weather''. 

As discussed below, the implicit function used in the level set method to represent the fire boundary is a time-varying signed distance function  that is referenced in space and time. Spatio-temporal data are ubiquitous in the environmental sciences and multi-level hierarchical representations have long been used to model such complex processes \citep[e.g., see][for general discussion]{cressie, banejree}. Likewise, the dependency observed in space or time has triggered intellectual curiosity in the deep learning community. \citet{zammit} provide a recent overview of statistical and AI approaches for deep learning spatial and spatio-temporal data.  Convolutional neural networks (CNNs), recurrent neural networks (RNNs), and their variants are the primary approaches in the deep learning literature for such data. In particular, hybrid convolutional recurrent networks \citep[e.g.,][]{zuo} have worked well for spatio-temporal data. For example,
in the context of wildfire,  \citet{burge} proposed the EPD-convLSTM model, which accounts for spatial and temporal dependency by  combining a convolutional long short-term memory (LSTM) RNN and an Encoder-Processor-Decoder (EPD).
However, these methods do not provide a mechanism for uncertainty quantification, do not directly incorporate physical relationships (such as implied by the level set method), and require a large amount of training data. 

An enhancement to traditional deep learning neural network approaches is to use so-called ``physics-informed neural networks'' that seek to encode physical relationships into the deep neural architecture \citep[e.g.,][]{raissi2019physics}. These methods have recently been considered for level-set fire front propagation in \citet{dabrowski2022bayesian}. Their work is implemented in a Bayesian framework that allows for data assimilation and uncertainty quantification. However, this approach does not consider realistic data-driven spatial and temporally varying velocities or large-scale megafire spread, which is our interest here. Alternatively, \citet{yoo} considered a hybrid physical/statistical level set model within a hierarchical Bayesian framework. Their approach included a latent vector autoregressive stochastic spatio-temporal process for velocity, and accommodates data assimilation and uncertainty quantification naturally in the Bayesian inference paradigm. However, as with many spatio-temporal Bayesian models, this method is fairly computationally intensive and is somewhat limited in that the latent dynamical process is assumed to be linear.  

We consider here an alternative approach for efficient physics-informed neural models that provides realistic data-driven estimates of fire front velocities as well as realistic uncertainty quantification. Specifically, we utilize a reservoir computing recurrent neural model known as an echo state network (ESN).  Hybrid ESN/statistical approaches for spatio-temporal data were introduced into statistics by \citet{mcdermott,McDermott2} and have subsequently been used in numerous applications \cite[e.g.,][]{bonas2021calibration,huang2021forecasting,barredo2022post}. Rather than estimate neural weight parameters, ESNs and other reservoir models draw weights randomly from a reservoir distribution, and only train the output layer parameters. This provides a distinct advantage over traditional neural models in that much less data is required to train ESNs, making them ideal for situations such as real-world fire spread where fire boundary data are only available sporadically (e.g., daily). As with most neural methods, two big challenges with ESNs relate to hyperparameter tuning \citep[e.g.,][]{wu,Probst} and uncertainty quantification \citep[e.g.,][]{McDermott2}.  

In this paper, we propose an ensemble ESN within the level set framework, which combines the mechanistically-motivated dynamic model (the level set method) with reservoir modeling (ESN) of velocities. There are several contributions of this work. 
Retaining the level set method as the physically-motivated dynamical core, our model enables a simple representation of any complicated boundary and its topological change through time. The ESN, nested within the level set method to learn propagation speeds, can improve forecasting performance by accommodating more complicated (i.e., nonlinear) dynamics that are present in many dynamic systems \citep{chen1992}. Importantly, the low computational cost of ESNs makes our model suitable and attractive for practical use. In that regard, our method addresses the neural hyperparameter tuning issue efficiently by randomly sampling the hyperparameters from plausible distributions to construct forecast ensembles. Finally, our method can quantify the uncertainty in forecasting by using the ensembles with novel calibration. Compared to forecast intervals derived from the Bayesian approach of \citet{yoo}, our method provides less formal forecast intervals, but they are effective and much less computationally expensive. We demonstrate the forecasting performance of the model on two simulated datasets generated from actual wind fields and two wildfires in California -- the Thomas and Haypress megafires.

The paper is organized as follows. Section \ref{sec:2} provides a brief background on ESNs and the level set method. Then, in Section \ref{sec:3}, our proposed hybrid level set/ESN model is described, with a novel calibration algorithm to quantify the uncertainty. We show the forecasting performance of our proposed model through simulation experiments and real data in Section \ref{sec:4} and \ref{sec:5}, respectively. Finally, Section \ref{sec:6} provides a brief discussion.

\section{Background}\label{sec:2}
This section provides brief background on ESNs and the level set method to facilitate discussion of the new model in Section \ref{sec:3}.

\subsection{Echo State Network (ESN)}\label{subsec:2.1}
The ESN has been an appealing type of recurrent neural network due to its relatively inexpensive computational cost, reduced training data burden, and conceptual simplicity \citep{lukosevicious}. The basic vanilla ESN model for time $t=1,...,T$ can be written as follows:
\begin{align} 
\label{eqn:eqn1}
\text{response}: &\quad \bm{v_t}=g_o(\bm{o}_t) \nonumber \\
\text{output}: &\quad \bm{o}_t= \bm{W}^{out}\bm{h}_t \nonumber \\
\text{hidden state}: &\quad \bm{h}_t= (1-\alpha_{\ell}) \bm{h}_{t-1} + \alpha_{\ell} \tilde{\bm{h}}_t \nonumber \\
&\quad \tilde{\bm{h}}_t =g_h\bigg(\frac{\nu}{|\lambda_w|}\bm{W}^{in}\bm{h}_{t-1} + \bm{U}\bm{x}_t\bigg), \\
\text{parameters}:  &\quad  \bm{W}^{in}=[w_{i,\ell}^{in}]_{i,\ell}:\gamma_{i,\ell}^{w}Unif(-a_w,a_w)+(1-\gamma_{i,\ell}^{w})\delta_0 \nonumber\\
&\quad \bm{U}=[u_{i,j}]_{i,j}:\gamma_{i,j}^{u}Unif(-a_u,a_u)+(1-\gamma_{i,j}^{u})\delta_0 \nonumber\\
&\quad \gamma_{i,\ell}^{w} \sim Bern(\pi_w) \nonumber\\
&\quad \gamma_{i,j}^{u} \sim Bern(\pi_u),\nonumber
\end{align}
where $N \times 1$ vector $\bm{v}_t$, $N \times 1$ vector $\bm{o}_t$, $J \times 1$ vector $\bm{h}_t$, and $n_u \times 1$ vector $\bm{x}_t$ correspond to the response, output, hidden units, and input signal, respectively. Critically, the $J \times J$ matrix $\bm{W}^{in}$ and $J \times n_u$ matrix $\bm{U}$ are pre-specified weight matrices with elements drawn randomly from specified distributions with hyperparameters $a_w$, $a_u$, $\pi_w$ and $\pi_u$. $\delta_0$ is a Dirac delta function at zero. Note that hyperparameters $\pi_w$ and $\pi_u$ control the sparsity of the matrices $\bm{W}^{in}$ and $\bm{U}$, respectively. Additionally, $\lambda_w$ is the spectral radius of $\bm{W}^{in}$, $\alpha_{\ell} \in (0,1]$ is the so-called ``leaking rate'', and $\nu$ is a tuning parameter. The element-wise activation functions $g_o(\cdot)$ and $g_h(\cdot)$ are user-specified, but $tanh(\cdot)$ functions are typically chosen for $g_h(\cdot)$, an identity function is typically chosen for $g_o(\cdot)$ when the response is continuous, and a softmax functions is typically chosen for categorical responses.  As discussed by \citet{lukosevicious}, the role of the hidden units in an ESN is to project the inputs into a higher dimensional space through nonlinear expansion and to simultaneously remember the input. 

To approximate the nonlinear system well, an ESN should exhibit the  ``echo state property''  \citep{zhang}, which essentially ensures that the reservoir model will eventually not be sensitive to initial conditions.  This is ensured in most cases if the spectral radius of $\bm{W}^{in}$ is less than 1 \citep{lukosevicious}. The ratio of the parameter $\nu$ and the largest eigenvalue $\lambda_w$ of $\bm{W}^{in}$ in the hidden state control the effective spectral radius. That is, $0 < \nu < 1$ guarantees the spectral radius of the model will be less than 1. 

The key to reservoir methods such as the ESN is that one only needs to estimate the $N \times J$ read-out matrix $\bm{W}^{out}$, which provides tremendous savings in terms of computational cost and the amount of data required to train the model. However, to prevent overfitting, it is very important that this estimation is regularized, which is typically accomplished using an $\ell_2$ penalty (i.e., ridge regression). For illustration, one can write the loss function, assuming the identity function for $g_o (\cdot)$, as
\begin{align}
\label{eqn:eqn2}
\widehat{\bm{W}}^{out}= \underset{\bm{V}}{\arg\min} \bigg[\frac{1}{N} \sum_{t=1}^{T} \bigg((\bm{v_t} -\bm{W}^{out} \bm{h_t})^{\top}(\bm{v_t} -\bm{W}^{out} \bm{h_t}) \bigg) + \tau \sum_{j=1}^{J}||\bm{w}^{out}_{\cdot j}||^2\bigg],
\end{align}
where $\bm{w}^{out}_{\cdot j}$ is $j^{th}$ column in $\bm{W}^{out}$, $\tau \in (0,\infty)$ is the regularization parameter, and $||\cdot||$ is the $\ell_2$ norm. One can easily obtain the solution for (\ref{eqn:eqn2}) in closed form \citep{lukosevicious}. 

Despite its conceptual simplicity and computational efficiency, selection of the tuning parameters in the ESN can be burdensome. \citet{lukosevicious} provides concise guidelines for the hyperparameters $\pi_w,\pi_u,a_w$, and $a_u$. In particular, they recommend sparse connections 
between units (i.e., small $\pi_w$ and $\pi_u$) to induce sparsity in $\bm{W}^{in}$ and $\bm{U}$, but recommend that $\bm{U}$ be more dense than $\bm{W}^{in}$. The role of $a_u$ and $a_w$ is to determine the effect of input $\bm{x_t}$ and previous hidden state $\bm{h_{t-1}}$ on the current hidden state $\bm{h_{t}}$ \citep{lukosevicious}. In addition to $\pi_w,\pi_u,a_w$, and $a_u$, one must select the reservoir size $J$, the leakage rate $\alpha_{\ell}$, and the echo-state parameter $\nu$. Selection of the hyperparameters $\alpha_{\ell}$ and $J$ can be accomplished by cross-validation or validation, but can be computationally inefficient.  \citet{mcdermott} used grid search with a validation approach for finding the best set of hyperparameters, and \citet{McDermott2} used a genetic algorithm \citep[e.g.,][]{Sivanandam}   in a ``deep'' ESN to accommodate the increased number of hyperparameters present in deep ESNs. \citet{Ribeiro} applied a Bayesian optimization algorithm to an ESN for short-term load forecasting, and \citet{Thiede} proposed a gradient-based optimization algorithm, which was shown to be superior to grid search for large numbers of hyperparameters. A systematic review of hyperparameter optimization for ESNs can be found in \citet{Matzner}.

Historically, the quantification of uncertainty in neural models presents a significant challenge \citep[e.g., see][for general discussion]{Abdar,zammit}. This is also true with reservoir methods. One useful approach that has been used for ESNs is to generate ensembles from the reservoir \citep{yao,sheng}. For example, \citet{mcdermott} and \citet{McDermott2} constructed forecasting ensembles by sampling the recurrent and input weight matrices independently from their reservoir distributions. In a Monte Carlo sense, the (1-$\alpha$)\% forecasting intervals were then obtained from the element-wise percentiles of the forecasting ensembles. \citet{bonas2021calibration} also generated ensemble forecasts on the training set and quantified the uncertainty by calibrating the residual with monotone cubic spline interpolation. In addition to the ensemble approach, \citet{McDermott2} adopted stochastic search variable selection priors \citep{George} for their Bayesian ESN. \citet{Atencia} employed Monte Carlo Dropout \citep{gal} as another straightforward strategy, which is similar to the ensemble approaches in principle. 
\subsection{The Level Set Method}\label{subsec:2.2}
The level set method (or the level set equation), proposed by \citet{osher2}, is an advection-diffusion equation used to model the evolution of an implicit function in an arbitrary dimension. That is,  
\begin{align}
\label{eqn:eqn3}
\frac{\partial \phi(\bm{s},t)}{\partial t}+ \bm{v}(\bm{s},t)\cdot \nabla \phi(\bm{s},t)=0,
\end{align}
where $\phi(\bm{s},t)$ is an implicit function evaluated at location $\bm{s}$ and time $t$, $\frac{\partial \phi(\bm{s},t)}{\partial t}$ denotes the temporal partial derivative, and $\bm{v}(\bm{s},t)=(v_x(\bm{s},t),v_y(\bm{s},t))^{\top} \in \mathbb{R}^2$ is the velocity that controls the direction and speed of the implicit function evolution. By construction, the level set method requires an implicit function, leading to the Eulerian representation of the boundary in wildfire applications. In Eulerian approaches, one represents the boundary implicitly through a contour function, as opposed to  Lagrangian frameworks where one defines a set of points on the boundary to describe the boundary explicitly. \citep{osher}. Eulerian representations can be computationally more expensive than Lagrangian methods, yet Eulerian approaches are typically more effective in representing complicated boundaries and their change through time; for example, when wildfires exhibit topological changes such as merging or splitting.
\begin{figure}[t]
\centering
\includegraphics[width=1\linewidth]{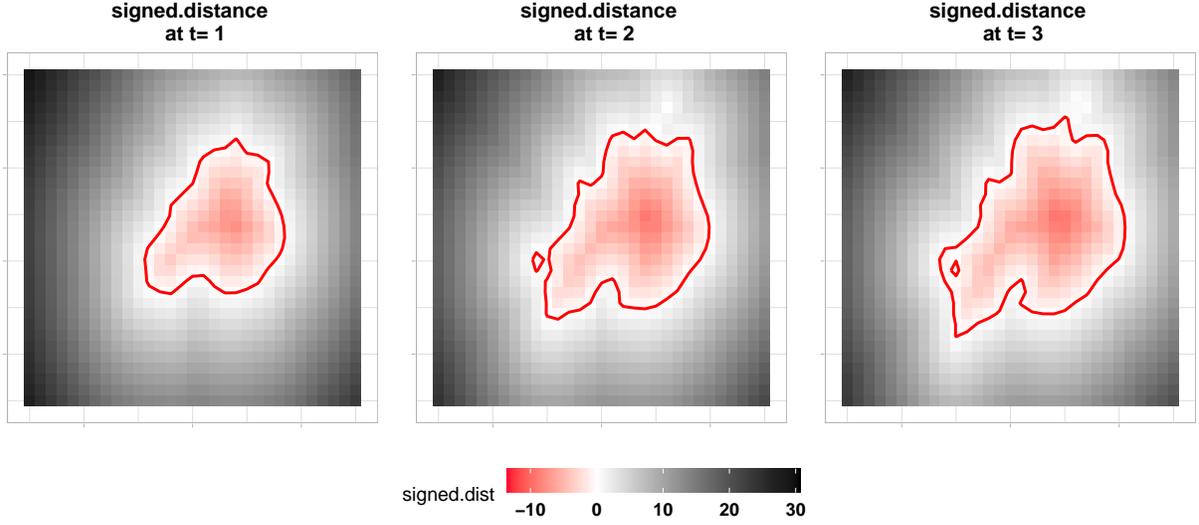}
  \caption{Illustration of a signed distance function: Three boundaries at $t=1,2,3$ are depicted from the left to  right.}
    \label{fig1}
\end{figure}

It is common to use a signed distance function for the implicit function in the level set method due to its straightforward interpretation and mathematical simplicity. The signed distance function $\phi(\bm{s})$ for $\forall \bm{s} \in \mathbb{R}^2$ is 
\begin{align}
\label{eqn:eqn4}
\phi(\bm{s})=\begin{cases}
-d(\bm{s}), &\quad \bm{s}\in \Omega^{-}\\
0, &\quad \bm{s}\in \partial\Omega\\
d(\bm{s}), &\quad \bm{s}\in \partial\Omega^{+}
\end{cases},
\end{align}
where $\Omega^{-}, \Omega^{+}$, and $\partial\Omega$ are the interior, and exterior of the boundary and the boundary itself, respectively. Here, $d(\bm{s})$ is the closest distance from $\bm{s}$ to the boundary. Assuming a distance given by the $\ell_2$ norm, 
\begin{align}
\label{eqn:eqn5}
d(\bm{s})=\min(||\bm{s}-\bm{s}_I||),
\end{align}
where $\bm{s}_I \in \partial \Omega$ is the set of points on the boundary. For an illustration of a signed distance function evolving in time, see Figure \ref{fig1}. During the evolution, one can see that a new boundary appears at $t=2$ on the bottom left, and a closed loop exists inside the main boundary at $t=3$. Regardless of any topological change the boundary exhibits, one can represent the boundary with the set $\partial \Omega=\{\bm{s}: \phi(\bm{s})=0\}$. Additionally, a signed distance function has the convenient mathematical property that
\begin{align}
\label{eqn:eqn6}
||\nabla \phi(\bm{s},t)||=1,
\end{align}
where $\nabla\phi(\bm{s},t)$ indicates a spatial gradient of $\phi(\bm{s},t)$. 

Notably, the frequently-made assumption that the evolution is in the normal direction to the boundary \citep[e.g.,][]{mallet, Alessandri, yoo} and (\ref{eqn:eqn6}) simplifies (\ref{eqn:eqn3}) so that
\begin{align}
\label{eqn:eqn7}
\frac{\partial \phi(\bm{s},t)}{\partial t}+ v_n(\bm{s},t)&=0,
\end{align} 
where $v_n(\bm{s},t)$ is the scalar velocity (speed) in the normal direction.  (\ref{eqn:eqn7}) can then be simply approximated by a forward Euler discretization
\begin{align}
\label{eqn:eqn8}
\frac{\phi(\bm{s},t+\triangle t)-\phi(\bm{s},t)}{\triangle t}+ v_n(\bm{s},t)&=0,
\end{align} 
where $\triangle t$ is time-step size.
Although one can still solve (\ref{eqn:eqn3}) without assumption (\ref{eqn:eqn6}) with numerical solvers \citep{osher}, it induces more computational cost. Importantly, there is not much to be gained by such complexity in data-driven approaches so long as one can learn the spatial and temporally varying speed in the normal direction, $v_n(\bm{s},t)$.
Indeed, as discussed in Section \ref{sec:1}, it is crucial in wildfire applications to estimate this to obtain realistic fire boundary propagation in complex fires. Building on the flexibility and convenience of the ESN, Section \ref{sec:3} presents a novel model that embeds an ESN within the level set equation to estimate $v_n (\bm{s},t)$ in (\ref{eqn:eqn8}).

\section{Hybrid Level Set/ESN Model}\label{sec:3}
This section presents a model in which an ESN, used to learn propagation speed, is embedded within the level set method. Section \ref{sec:3.1} illustrates the structure of the model. Section \ref{sec:3.2}, \ref{sec:3.3}, and \ref{sec:3.4} present the hyperparameter tuning approach, uncertainty quantification method, tuning selection, and the algorithm, respectively.

\subsection{Nested ESN Model}\label{sec:3.1}
The hybrid level set/ESN model considers a vectorized form of equation (\ref{eqn:eqn8}) with additive error in which the spatio-temporal speeds in the normal direction to the boundary are specified via an ESN. Specifically, for spatial locations $\bm{s_i}, 1\leq i \leq N$ in the domain of interest $\mathcal{D} \subset \mathbb{R}^2$ and time indexed by $t$, $1\leq t\leq T$, the full model is given by:
\begin{align}
\label{eqn:eqn9}
\text{response}: &\quad \bm{\phi}_t=\bm{\phi}_{t-\triangle t}-\bm{v}_{n,t-\triangle t} \triangle t+ \bm{\epsilon}_{t} \nonumber \\
\text{rate of spread}: &\quad \bm{v}_{n,t-\triangle t}= \bm{W}^{out} \bm{h}_{t-\triangle t} \nonumber\\
\text{hidden state}: &\quad \bm{h}_{t-\triangle t}= (1-\alpha_{\ell}) \bm{h}_{t-2\triangle t} + \alpha_{\ell} \tilde{\bm{h}}_{t-\triangle t} \nonumber \\
&\quad \tilde{\bm{h}}_{t-\triangle t} =g_h\bigg(\frac{\nu}{|\lambda_w|}\bm{W}^{in}\bm{h}_{t-2 \triangle t} + \bm{U}\bm{x}_{t-\triangle t}\bigg) \\
\text{parameters}: &\quad  \bm{W}^{in}=[w_{i,\ell}^{in}]_{i,\ell}:\gamma_{i,\ell}^{w}Unif(-a_w,a_w)+(1-\gamma_{i,\ell}^{w})\delta_0 \nonumber\\
&\quad \bm{U}=[u_{i,j}]_{i,j}:\gamma_{i,j}^{u}Unif(-a_u,a_u)+(1-\gamma_{i,j}^{u})\delta_0 \nonumber\\
&\quad \gamma_{i,\ell}^{w} \sim Bern(\pi_w) \nonumber\\
&\quad \gamma_{i,j}^{u} \sim Bern(\pi_u),\nonumber
\end{align}
where $\bm{\phi}_t = (\phi(\bm{s_1},t),\ldots,\phi(\bm{s_N},t))^{\top}$, $\bm{v}_{n,t-\triangle t}= (v_n(\bm{s_1},t-\triangle t),\ldots,v_n(\bm{s_N},t-\triangle t))^{\top}$, and $\bm{\epsilon}_{t}=(\epsilon(\bm{s_1},t),\ldots, \epsilon(\bm{s_N},t))^{\top}$ denote an observed signed distance function, rate of spread in the normal direction to the boundary, and error term on the space $\mathcal{D}$ at the indexed time, respectively. 
Note that rate of spread (speed), hidden state, and parameters model are equivalently defined as in (\ref{eqn:eqn1}). 

The nested ESN model used to estimate the rate of spread $\bm{v}_{n,t-\triangle t}$ can provide more data-driven learning flexibility than the linear formulation in \cite{yoo} and the Rothermel model \citep{rothermel}. Importantly, one only needs to estimate the read-out matrix $\bm{W}^{out}$ after the appropriate hyperparameters are specified. If ridge regression is used as in (\ref{eqn:eqn2}), one also must specify the penalty coefficient $\tau$, but this is still computationally efficient. 

As indicated in Section \ref{sec:2}, uncertainty quantification and the selection of hyperparameters in this model must be accounted for.  We quantify forecast uncertainty through ensembles, but our method is motivated by the highest density region proposed by \citet{Hyndman} and calibration strategy of \citet{ranadeep} as described in Section \ref{sec:3.2}. We adopt a random search approach for hyperparameter tuning \citep{Bergstra} as described in Section \ref{sec:3.3}. The complete algorithm for our hybrid models is presented in Section \ref{sec:3.4}.

\subsection{Uncertainty Quantification}\label{sec:3.2}
Once the model is fitted, a one-step ahead forecast for time $T+1$ is given by 
\begin{align}
\label{eqn:eqn10}
\widehat{\bm{\phi}}_{T+1} = \bm{\phi}_{T} -\widehat{\bm{W}}^{out} \bm{h}_{T} \triangle t,
\end{align}
where $\bm{h}_{T}$ is obtained from (\ref{eqn:eqn9}) given $\bm{W}^{in}$, $\bm{U}$, $\alpha_{\ell}$, $J$ and $\nu$. To quantify the uncertainty associated with this forecast, we take the ensemble approach motivated by \citet{sheng},\citet{yao}, and \citet{mcdermott}. That is, we collect an ensemble of $N_{ensemble}$ forecasts $\widehat{\bm{\phi}}_{en,T+1}=(\widehat{\bm{\phi}}_{T+1}^{(1)},\ldots,\widehat{\bm{\phi}}_{T+1}^{(N_{ensemble})})$ where each forecast is associated with different hyperparameters and reservoir matrices. We then construct $(1-\alpha)$\% forecast intervals at location $\bm{s}$ as 
\begin{align}
\label{eqn:eqn11}
\bigg( \hat{\ell}_{T+1}(\bm{s}),\hat{u}_{T+1}(\bm{s})\bigg)=\underset{\ell_{T+1}(\bm{s}),u_{T+1}(\bm{s})}{\arg\min}
\bigg( u_{T+1}(\bm{s})-\ell_{T+1}(\bm{s})\bigg)
\end{align}
such that 
\begin{align}
\label{eqn:eqn12}
\frac{1}{N_{ensemble}}\sum_{i=1}^{N_{ensemble}}I\bigg(\ell_{T+1}(\bm{s}) \leq \widehat{\phi}^{(i)}_{T+1}(\bm{s})\leq u_{T+1}(\bm{s}) \bigg)= 1-\alpha ,
\end{align}
where $\ell_{T+1}(\bm{s})$ and $u_{T+1}(\bm{s})$ are the interval lower and upper bounds.

The motivation for (\ref{eqn:eqn12}) is from the calibration strategy of \citet{ranadeep}. To quantify the uncertainty for spatial prediction in reservoir models, \citet{ranadeep} first obtained the median and inter-quartile range from ensembles and trained the optimal calibration cutoff value to ensure a $(1-\alpha)$\% coverage rate over the training set. Then, the prediction intervals for their test set were constructed from the median and interquartile range from the ensembles over the test set and the cutoff value obtained from the training set. Here, we do not require a separate training set calibration because the coverage rate is given directly in (\ref{eqn:eqn12}) based on ensembles of forecasts.

The highest density region proposed by \citet{Hyndman} motivates the relationship in (\ref{eqn:eqn11}). The highest density region provides the shortest intervals, which may consist of several disjoint intervals for multimodal distributions \citep{Hyndman}. However, several disjoint intervals complicate the interpretation in wildfire applications, which leads to quantifying the uncertainty by single intervals as given in (\ref{eqn:eqn11}). The aim of using the (1-$\alpha$)\% highest density intervals is to exclude the poor forecasts (e.g., outliers), which we find effective in our simulation experiments and real data analysis.

\subsection{Hyperparameter tuning }\label{sec:3.3}
Hyperparameter selection is one of the most important components when building deep neural models \citep{Probst}. The most intuitive and straightforward approach is to use a grid search to explore the parameter space. However, such methods are inefficient for  relatively high-dimensional parameter spaces. An efficient alternative is to use random search  \citep[e.g.,][]{Bergstra}, which improves on grid search by randomly searching over hyperparameters from specified ``prior'' distributions \citep{yu}, and can be implemented at a fraction of the cost of optimization algorithms.  
For an extensive review of hyperparameter optimization strategies, refer to \citet{yu}.

We use a version of random search for our hybrid level set/ESN algorithm. The ensemble approach to ESN modeling allows one to not only sample from the reservoirs for each ensemble member, but also the hyperparameter distributions.  Specifically, we draw $a_u$, $a_w$, $J$, $\tau$, $\nu$, and $\alpha_{\ell}$ for each ensemble member independently from their prior distributions.  The guidelines from \citet{lukosevicious} can help determine fixed values for $\pi_w$, $\pi_u$, and the distributions from which hyperparameters are drawn. 
It is important to note that significant computational efficiency is gained from selecting the hyperparameters through random search, and also provides uncertainty quantification as described in Section \ref{sec:3.2}.

\subsection{Hybrid level set/ESN algorithm}\label{sec:3.4}

\begin{algorithm}
    \caption{Hybrid Level Set/ESN Algorithm}
    \label{algo:1}
    \begin{algorithmic}[1]
        \State \textbf{Input} Data $\{\bm{\phi_t}\}$ for $t=1,...,T$. The prior distributions of $a_u$, $a_w$, $J$, $\tau$, $\nu$, and $\alpha_{\ell}$ denoted by $P(a_u)$, $P(a_w)$, $P(J)$, $P(\tau)$, $P(\nu)$, $P(\alpha_{\ell})$, respectively. Fixed constants $\pi_w$ and $\pi_u$. 
        \State \textbf{for} $k=1:N_{ensemble}$ \textbf{do}
        \State \hskip1.0em Randomly draw $a_u^{(k)}$, $a_w^{(k)}$, $J^{(k)}$, $\tau^{(k)}$, $\nu^{(k)}$, and $\alpha_{\ell}^{(k)}$ from prior distributions. \label{algo:step3}
        \State \hskip1.0em Construct $\bm{W}^{in,(k)}$, $\bm{U}^{(k)}$ with sampled $a_w^{(k)}$, $a_u^{(k)}$, $\nu^{(k)}$, $\pi_w$, and $\pi_u$. \label{algo:step4}
        \State \hskip1.0em Fit the model in (\ref{eqn:eqn9}) with $J^{(k)}$ and $\alpha_{\ell}^{(k)}$, and estimate $\widehat{\bm{W}}^{out,(k)}$ by ridge regression with $\lambda^{(k)}$.\label{algo:step5}
        \State \hskip1.0em Obtain the forecasts $\widehat{\bm{\phi}}_{T+1}^{(k)}$ using $\widehat{\bm{W}}^{out,(k)}$.\label{algo:step6}
        \State \textbf{end for}
        \State Define the ensemble of forecasts: $\widehat{\bm{\phi}}_{en,T+1}=\{\widehat{\bm{\phi}}_{T+1}^{(1)},\ldots,\widehat{\bm{\phi}}_{T+1}^{(N_{ensemble})}\}$
        \State Ensemble forecast median: $\widehat{\bm{\phi}}_{med,T+1}$=median($\widehat{\bm{\phi}}_{en,T+1}$)
        \State Ensemble uncertainty: ($\hat{\ell}_{T+1}$, $\hat{u}_{T+1}$) obtained by (\ref{eqn:eqn11}) and (\ref{eqn:eqn12}) corresponding to the significance level $\alpha$.
        \State \textbf{output} $\widehat{\bm{\phi}}_{med, t+1}$ and ($\hat{\ell}_{T+1}$, $\hat{u}_{T+1}$)
    \end{algorithmic}
\end{algorithm}
The hybrid level set/ESN model, including the uncertainty quantification and hyperparameter selection described in Section \ref{sec:3.2} and \ref{sec:3.3}, is detailed in Algorithm \ref{algo:1}. 

Step \ref{algo:step3} of Algorithm \ref{algo:1} draws hyperparameters $a_u^{(k)}$, $a_w^{(k)}$, $J^{(k)}$, $\tau^{(k)}$, $\nu^{(k)}$, and $\alpha_{\ell}^{(k)}$ from their prior distributions (i.e., random search), from which recurrent and input weight matrices $\bm{W}^{in,(k)}$ and $\bm{U}^{(k)}$ are constructed in Step \ref{algo:step4}. Fitting model (\ref{eqn:eqn9}) and estimating $\widehat{\bm{W}}^{out,(k)}$ with $\tau^{(k)}$ is conducted in Step \ref{algo:step5}, and Step \ref{algo:step6} produces forecast ensemble members  
\begin{align}
\label{eqn:13}
\widehat{\bm{\phi}}_{T+1}^{(k)} = \bm{\phi}_{T} -\widehat{\bm{W}}^{out,(k)} \bm{h}_{T}^{(k)} \triangle t,
\end{align}
where $\bm{h}_{T}^{(k)}$ is constructed by (\ref{eqn:eqn9}) with $\bm{W}^{in,(k)}$, $\bm{U}^{(k)}$, $\alpha_{\ell}^{(k)}$, $J^{(k)}$ and $\nu^{(k)}$. After $N_{ensemble}$ iterations of Steps \ref{algo:step3} to \ref{algo:step6}, which creates the ensembles $\widehat{\bm{\phi}}_{en,T+1}=(\widehat{\bm{\phi}}_{T+1}^{(1)},\ldots,\widehat{\bm{\phi}}_{T+1}^{(N_{ensemble})})$, Algorithm \ref{algo:1} provides the median of $\widehat{\bm{\phi}}_{en,T+1}$ for the point forecast estimate and quantifies the uncertainty with $(1-\alpha)$\% intervals of $\widehat{\bm{\phi}}_{en,T+1}$ using (\ref{eqn:eqn11}) and (\ref{eqn:eqn12}). 

We show the forecasting performance of this algorithm through simulation experiments and with wildfire data in Sections \ref{sec:4} and \ref{sec:5}, respectively.
\begin{figure}[t]
\centering
\includegraphics[width=1\linewidth]{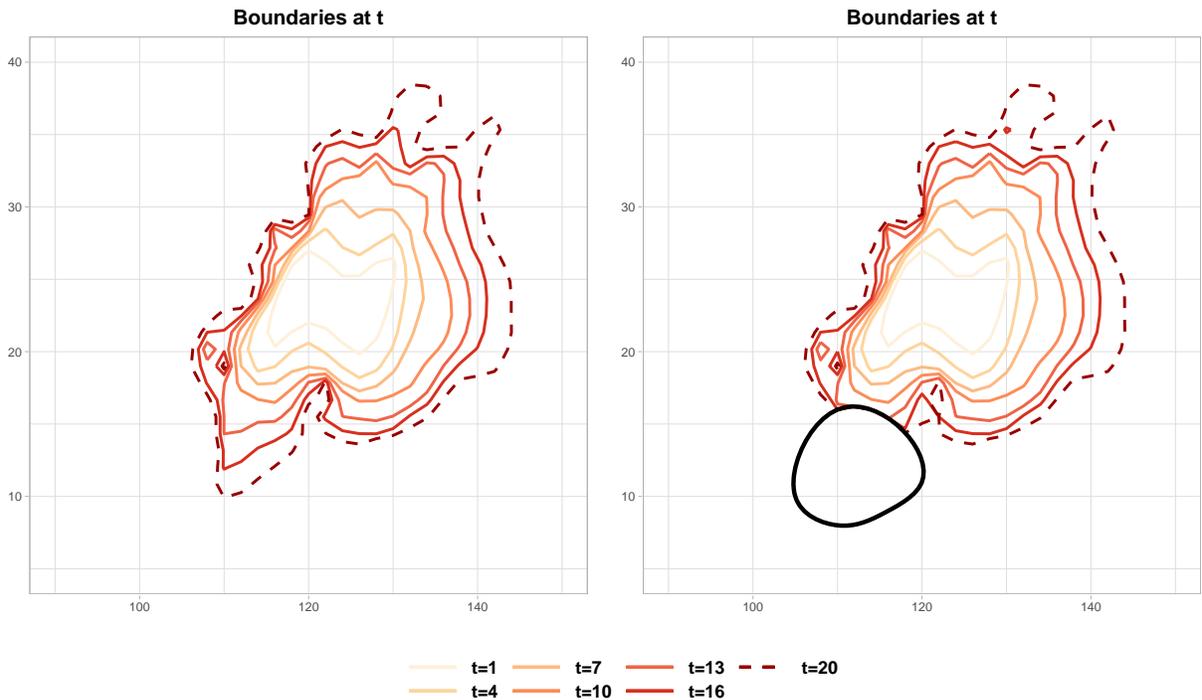}
  \caption{Simulation experiments: the evolution of boundaries for each simulation. In the left figure, the evolution to the northeast is prominent. In the right figure, evolution is equivalent to the first simulation, but constraint (solid black line) prevents the spread over it.}
    \label{fig2}
\end{figure}
\begin{figure}[t]
\centering
\includegraphics[width=0.95\linewidth]{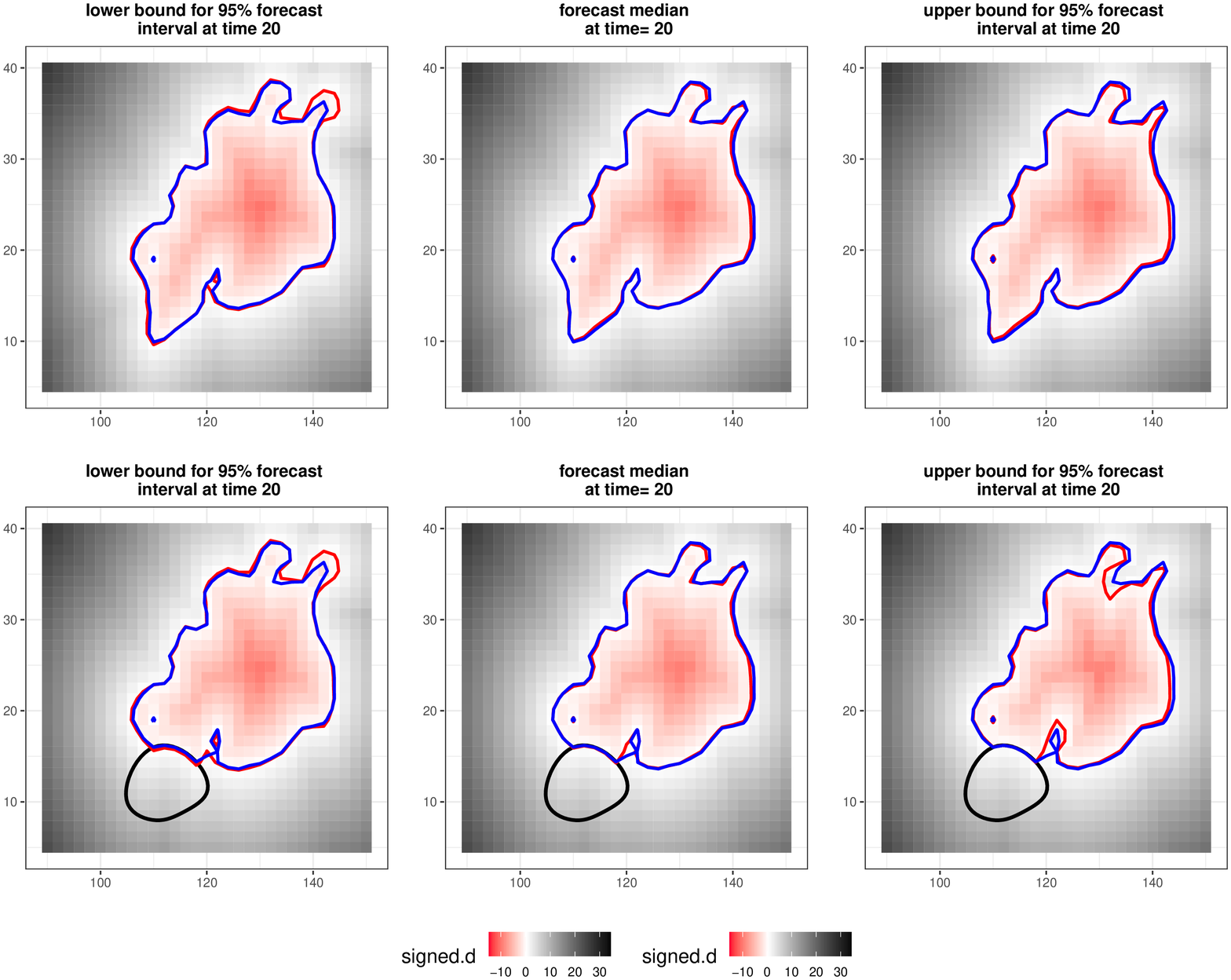}
  \caption{Simulation experiments: 95\% forecast intervals and forecast median. Figures in the first and second rows correspond to the first and second simulations. The solid blue line denotes the true boundary, and the solid red line indicates the predicted boundary.}
    \label{fig3}
\end{figure}
\section{Simulation experiments}\label{sec:4}
To make simulations more realistic, we generate two datasets by evolving a randomly generated initial closed boundary with the actual wind speeds obtained from the R package \emph{rWind} \citep{rwind}. Specifically, we first obtain twenty series of wind speeds from 2018-06-25 00:00:00 to 2018-07-04 12:00:00 at every twelve hours over longitude and latitude [90,150]$\times$[5,40]. Then, the initial closed boundary is evolved from these wind speeds using equation \ref{eqn:eqn8}. The first generated dataset (see the left panel in Figure \ref{fig2}) exhibits an evolution that includes disjoint boundaries merging into one. The second dataset (see the right panel of Figure \ref{fig2}) is generated with topological constraints, similar to a spatial area that might prevent wildfire spread (e.g., rivers, lakes, or oceans). Topological constraints are depicted in a solid black line for the second dataset. On both figures, the dashed line indicates the boundary we seek to forecast. 

We first obtain a signed distance function on a 31 $\times$ 31 grid on the domain $[90,150] \times [5,40]$. Then, we leave out data at $t=20$ to evaluate forecasting performance. That is, we fit our proposed model (\ref{eqn:eqn9}) on data from $t=1,\ldots,19$ and obtain a one-step ahead forecast at $t=20$ by Algorithm \ref{algo:1} with $\triangle t=0.1$. For the hyperparameter distributions, we set $P(a_u)=P(a_w)=(0.1,0.5,1)$, $P(\alpha_{\ell})=(0.01,0.0621,\ldots,0.9478,1)$, $P(\nu)=(0.1,0.2,\ldots,0.8,0.9)$, $P(\tau)=(0.001,1.112,\ldots,8.889,10)$, and $P(J)=(50,$ $\ldots,100)$. The values of $P(J)$ are relatively small to prevent overfitting, so each ensemble member functions as a relatively weak learner \citep{mcdermott}. Finally, we fix $\pi_w=0.3$ and $\pi_u=0.5$ based on the guidelines in \citet{lukosevicious}.

The choice of input is application-specific. \citet{mcdermott} used embeddings (lagged values of the data) for the input $\bm{x}_{t}$, which we follow. Specifically, we set
\begin{align}
\label{eqn:14}
\bm{x}_{t}=[\bm{\phi}_{t}^{\top},\bm{\phi}_{t-\triangle t}^{\top}]^{\top}.
\end{align}
Embeddings are motivated by a nonlinear Hamilton-Jacobi equation in the level set method. Namely, the level set method in (\ref{eqn:eqn7}) becomes a nonlinear Hamilton-Jacobi equation if $v_n(\bm{s},t)$ depends on $\phi(\bm{s},t)$ \citep{osher2}, which occurs in the ESN if we consider embeddings as in (\ref{eqn:14}). Finally, we set the initial value of the hidden state $\bm{h}_1=\bm{U} \times \bm{x}_{1+\triangle t}=\bm{U}\times [\bm{\phi}_{1+\triangle t}^{\top},\bm{\phi}_{1}^{\top}]^{\top}$.

Figure \ref{fig3} shows the 95\% forecast intervals and median from 3,000 ensembles. One can see that the 95\% forecast intervals are quite narrow, but they mainly capture the true boundary. The northeast part of the boundary shows the most variability, which aligns with the evolution pattern depicted in Figure \ref{fig2}. Moreover, the 95\% forecast intervals for the second simulation account for the topological constraints (although the predicted boundary spreads very slightly over the topological constraints in the lower bound). 

We adopt the threat score (TS) as a summary measure (e.g.,\citet{wikle2019}) to assess the quality of the forecast median.  Specifically, the TS is defined as
\begin{align}
\label{eqn:15}
TS=\frac{A_{11}}{A_{11}+A_{10}+A_{01}},
\end{align}
where $A_{11}$ is the common area where the model predicts an event that eventually occurs, $A_{10}$ is the area where the model expects an event to occur but not eventually occur, and  $A_{01}$ is the area where an event occurs, but the model fails to predict its occurrence. The range of TS summary measure is from 0 to 1, where a value of 1 indicates a perfect forecast. For a continuous variable such as a signed distance function, one needs a threshold parameter $\tau$ to determine the occurrence of an event. Therefore, we set $\tau=0$ since points on the boundary have a signed distance of 0. That is, if $\phi(\bm{s},t) \leq 0$, we consider that an event occurred on $\bm{s}$ at $t$. The TS summary measures for the forecast median of the first and second simulations are 0.969 and 0.963, demonstrating that our model provides  good forecasting performance.

\section{Wildfire applications}\label{sec:5}
\begin{figure}[t]
\centering
\includegraphics[width=1\linewidth]{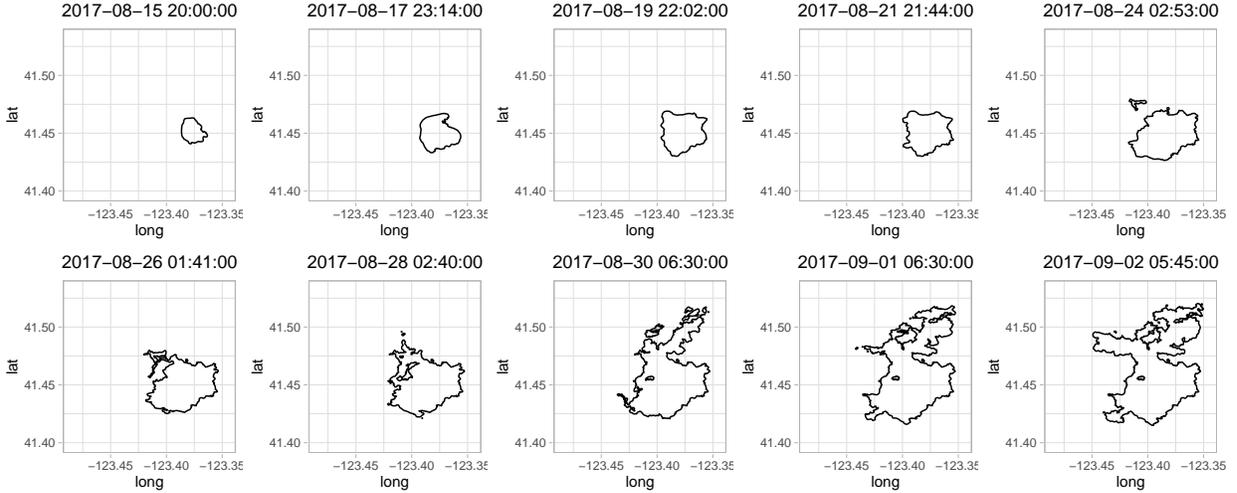}
  \caption{Fire boundaries for various times during the Haypress fire. Although the rate of spread was relatively slow until 2017-18-28 02:40:00, it subsequently accelerated, especially to the west at 2017-09-02 05:45:00.}
    \label{fig4}
\end{figure}

\begin{figure}[t]
\centering
\includegraphics[width=1\linewidth]{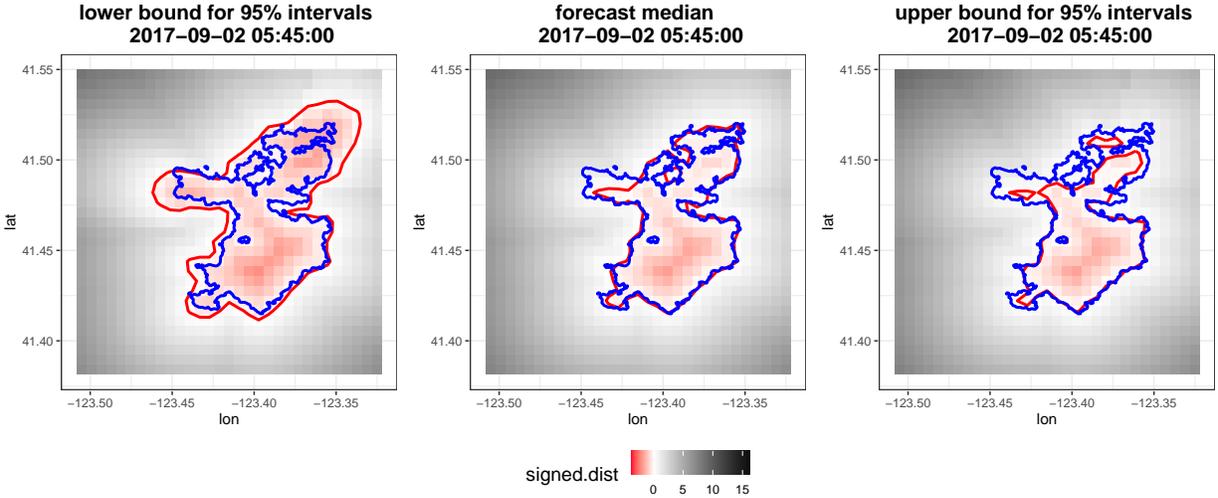}
  \caption{Forecast medians and 95\% intervals for the Haypress fire. The solid blue lines denote the true boundary, while the solid red lines indicates the predicted boundary. Most of the variability in forecast intervals is observed on the left part of the boundary.}
    \label{fig5}
\end{figure}

In this section we investigate the forecasting performance of our model on two real datasets - the Haypress and Thomas megafires that occurred in California in 2017. 
\subsection{The Haypress fire}
The Haypress fire, ignited in July 2017 and was contained in January 2018. It was the largest wildfire among the Orleans Complex fires, which burned over 27,000 acres in Siskiyou County, California. We consider the first 11 observations of the Haypress fire boundary from GeoMac database \citep{geomac} for this analysis. Specifically, we fit the model (\ref{eqn:eqn9}) on the first ten observations, leaving out the eleventh observation to evaluate forecasting performance. Figure \ref{fig4} shows the wildfire propagation from the second observation to the eleventh observation, from which one can see the rapid rate of spread to the north at 2017-08-30 06:30:00 and to the west at 2017-09-02 05:45:00. The forecasting task is more challenging than the simulation experiments in that a predominant expansion to the west shown at 2017-09-02 05:45:00 is not observed from 2017-08-13 23:32:00 to 2017-09-01 06:30:00. At 2017-09-01 06:30:00, the Haypress fire shows a slight spread to the west, but it is not as striking as observed in the boundary at 2017-09-02 05:45:00.

With a signed distance function on a $31 \times  31$ grid over [-123.5053,-123.3252] $\times$  [41.3840, 41.5471] by (\ref{eqn:eqn4}), we use Algorithm \ref{algo:1} to fit the model (\ref{eqn:eqn9}) with the same hyperparameter distributions and $\triangle t=0.1$ as in the simulation experiments. Also, the inputs and initial value of the hidden state are chosen in the same manner. Figure \ref{fig5} shows the 95\% forecast intervals and median from 3,000 ensembles. The forecast median captures the evolution to the west, and the 95\% intervals provide good coverage of the true boundary. Given the small number of observations used to train the model, the model visually shows very good forecasting performance. 

To more objectively compare the forecast performance of our model to that by \citet{yoo}, we measure the quality of forecast intervals using the interval score (IS) \citep{gneiting2}, which is given by
\begin{align}
\label{eqn:16}
S_{\alpha}^{Int}&(\ell(\bm{s},t),u(\bm{s},t),\phi(\bm{s},t))= \nonumber \\
&(u(\bm{s},t)-\ell(\bm{s},t))+\frac{2}{\alpha}(\ell(\bm{s},t)-\phi(\bm{s},t))I(\phi(\bm{s},t) <\ell(\bm{s},t)) \\
&+\frac{2}{\alpha}(\phi(\bm{s},t)-u(\bm{s},t))I(\phi(\bm{s},t) >u(\bm{s},t)) \nonumber,
\end{align}
where $\ell(\bm{s},t)$, $u(\bm{s},t)$, $\alpha$, and $\phi(\bm{s},t)$ are the lower bound, upper bound, significance level, and true signed distance function on $\bm{s}$ at $t$. The IS consists of the length of an interval $(u(\bm{s},t)-\ell(\bm{s},t))$ and the penalty $(\frac{2}{\alpha}(\ell(\bm{s},t)-\phi(\bm{s},t))$ or $ \frac{2}{\alpha}(\phi(\bm{s},t)-u(\bm{s},t)))$ induced if forecast intervals fail to contain $\phi(\bm{s},t)$. A smaller IS indicates better forecast intervals. The IS is intuitively appealing since it favors narrow forecast intervals but penalizes if those are too narrow. Furthermore, the IS is a proper scoring rule \citep{gneiting2}. We summarize the quality of forecasting intervals by 
\begin{align}
\label{eqn:17}
IS= \frac{\sum_{\bm{s}:\phi(\bm{s},t)\leq 0} S_{\alpha}^{Int}(\ell(\bm{s},t),u(\bm{s},t),\phi(\bm{s},t))}{  | \{\bm{s}:\phi(\bm{s},t)\leq 0\} |},
\end{align}
where $|\cdot |$ denotes the cardinality of the set and $t$=2017-09-02 05:45:00 for the Haypress fire. We only consider the locations inside the true boundary at 2017-09-02 05:45:00 since our primary interest is the boundary (indirectly, the burned and burning area), not the whole domain.

As given in Table \ref{table1}, the TS summary measure and IS from model applied to the Haypress fire forcast are 0.849 and 0.776, respectively, while those from \citet{yoo} are 0.791 and 1.730, demonstrating better forecast performance of the hybrid level set/ESN model. We also note that the computational cost of the hybrid level set/ESN is much lower than \citet{yoo}. Specifically, the algorithm presented in this paper takes less than 3 minutes on a standard laptop computer compared to 3 hours for the model in \citet{yoo} on a high-performance cluster. 
\begin{table}[t]
\centering
\caption{Comparison of Interval Score (IS) and Threat Score (TS) between models. $\mathcal{M}_1$ and $\mathcal{M}_2$ refer to the model presented here and the model by \citet{yoo}, respectively.}
\label{table1}
\begin{tabular}{ccccccc}
\hline
\multirow{2}{*}{Model} &  & \multicolumn{2}{c}{The Haypress fire} &  & \multicolumn{2}{c}{The Thomas fire} \\ \cline{3-4} \cline{6-7} 
                       &  & Interval Score     & Threat Score     &  & Interval Score    & Threat Score    \\ \hline
$\mathcal{M}_1$        &  & 0.776              & 0.849            &  & 4.901             & 0.964           \\
$\mathcal{M}_2$        &  & 1.730              & 0.791            &  & 5.286             & 0.954           \\ \hline
\end{tabular}
\end{table}
\subsection{The Thomas fire}
\begin{figure}[t]
\centering
\includegraphics[width=0.95\linewidth]{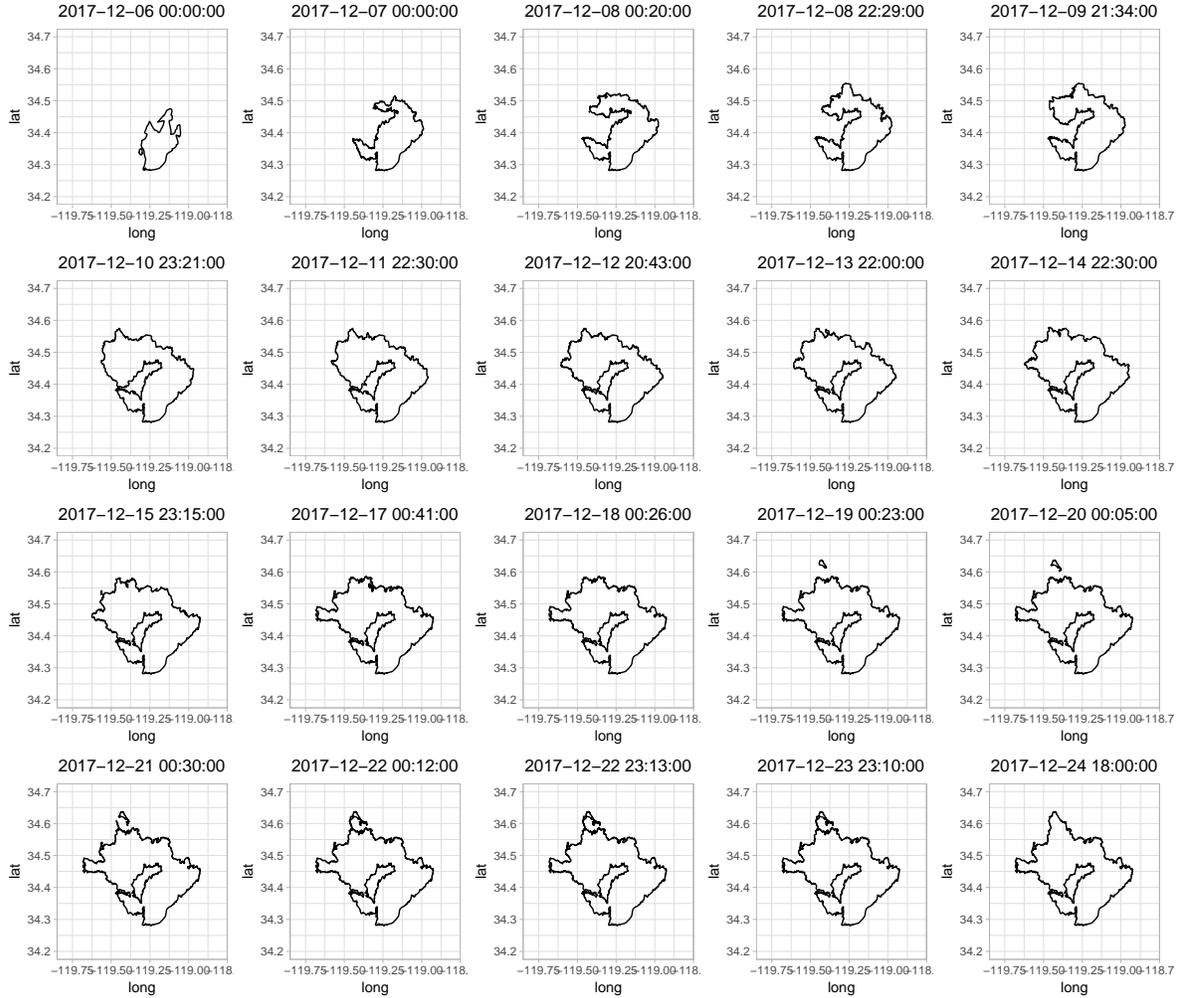}
  \caption{Thomas fire boundaries for 20 times Thomas fire (note, the first observation at 2017-12-05 04:40:00 is omitted). Compared to the Haypress fire, the rate of spread is relatively consistent in general.}
    \label{fig6}
\end{figure}
The massive Thomas fire in Southern California burned 114,000 hectares from the time of its ignition on December 4, 2017 to its containment in January 2018. Even after containment, post-fire debris flow continued to present hazards \citep{addison}. We consider 21 observations of fire boundaries from the GeoMac database \citep{geomac} as depicted in Figure \ref{fig6}. Notably, the spread rate is relatively more uniform than the Haypress fire, especially after 2017-12-10 23:21:00. 
However, there was a sudden extension of the boundary at 2017-12-24 18:00:00. This makes forecasting at 2017-12-24 18:00:00 challenging as the boundaries between 2017-12-19 00:23:00 and 2017-12-23 23:10:00 were evolving quite slowly.  

\begin{figure}
\centering
\includegraphics[width=0.95\linewidth]{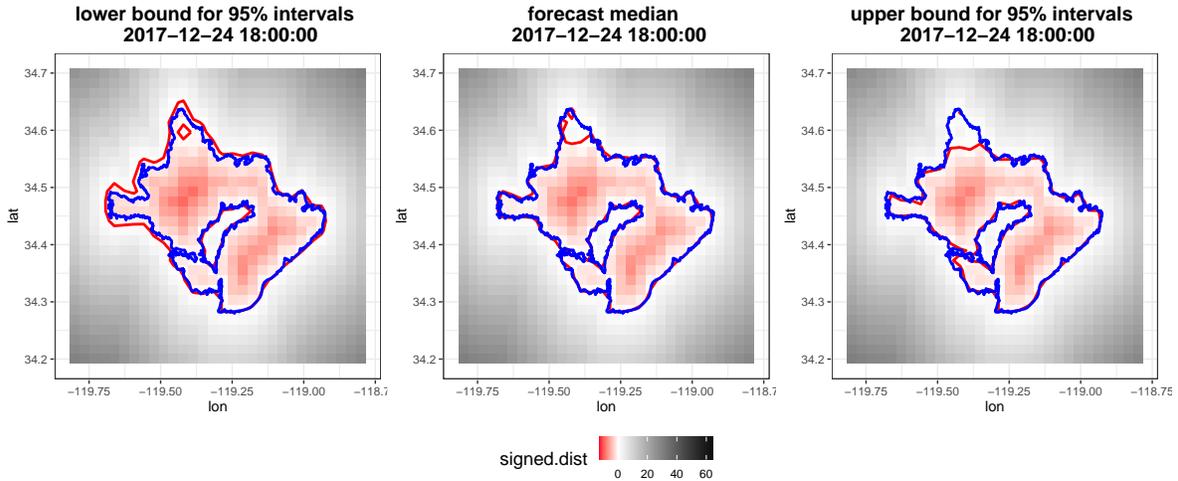}
  \caption{Forecast median and 95\% intervals for the Thomas fire. The solid blue line denotes the true boundary and the  solid red line indicates predicted boundary}
    \label{fig7}
\end{figure}

We use the first twenty observations to fit the model and evaluate forecast performance at 2017-12:24 18:00:00. A signed distance function is constructed on a $30 \times  30$ grid over [-119.8,-118.8] $\times$  [34.2, 34.7] by (\ref{eqn:eqn4}). Using the same hyperparameter distributions, inputs, $\triangle t=0.1$, and initial value of the hidden state as the simulation experiments, we obtain the forecast median and 95\% forecast intervals from 3,000 ensembles, as shown in Figure \ref{fig7}.  The forecast median fails to predict the upper part of the boundary connected at 2017-12-24 18:00:00, but importantly, the 95\% forecast intervals capture this feature with a loop in the lower bound. This is a significant benefit of uncertainty quantification approach presented here. 

We also compare objectively the hybrid level set/ESN model to the approach in \citet{yoo} with the TS summary measure and IS. As given in Table \ref{table1}, the hybrid model gives TS and IS values of 0.964 and 4.901, respectively, while \citet{yoo} gave 0.954 and 5.286. These summary measures demonstrate superiority of the hybrid model, but the difference is not as dramatic as with the Haypress fire. In addition, we observe visually that the 95\% credible intervals from \citet{yoo} cover the true boundary without a closed loop in the lower bound, which is found in Figure \ref{fig7}. Nevertheless, the wider credible intervals by \citet{yoo} have more negative impacts on the IS than the penalty induced by not fully covering the true boundary in the hybrid model.

\section{Discussion}\label{sec:6}
Although the level set method with a signed distance function has become a powerful tool in wildfire applications, the importance of data-driven learning is often neglected. Given the inhomogeneous environmental conditions and rapid spread of wildfires, it is important that the forecasting model adapt through data-driven learning. In addition, an important aspect of a trustworthy forecasting model is realistic uncertainty quantification, which is not available with deterministic model applications. 
Finally, a real-world wildfire forecasting model should account for nonlinearity in the fire evolution to accommodate more realistic dynamics.

We formulate our model by combining the mechanistically-motivated dynamic model (the level set method) with reservoir modeling of fire front propagation speed (ESN). Importantly, this model allows for data-driven learning as well as accounting for nonlinearity. Also, our algorithm is highly efficient and provides realistic forecasting intervals without requiring significant training costs and large amounts of data. Through simulation experiments, we demonstrate outstanding forecasting performance. We also show that our model can provide better forecasting intervals than a state-of-the-art Bayesian level set model \citep{yoo} on the Haypress and Thomas fire.

Note that we used the same distributions for hyperparameters for the simulations and real-world examples. Although for certain applications one might need to choose different distributions depending on the scale of data, our proposed random search algorithm is computationally efficient and relieves much of the burden in the tuning parameter selection as well as providing realistic forecasting intervals. Also, note that the same time-step size $\triangle t$ is used for both numerical experiments and wildfire applications. We observed that the results are not very sensitive to the choice of $\triangle t$ as the estimated readout matrix $\widehat{\bm{W}}^{out}$ adjusts to different values of $\triangle t$. Note, however, that our hybrid level set model is not limited to wildfire applications. The methodology presented here can also be used to infer shrinking boundaries, such as the response to a tumor after treatment, or the retreat of Antarctic glaciers. The key is the data-driven estimation of the speed in the normal direction to the boundary, which can easily accommodate expansion or contraction (or both in different parts of the domain). 

Our model shows superior results and is more computational efficient compared to \citet{yoo}, but this does not necessarily mean that our model is more advantageous than \citet{yoo} in every respect. Specifically, one limitation of our model is the failure to explicitly determine the effects of covariates (e.g., slope and landcover). Note that covariates, such as slope, aspect, existing vegetation cover, and forest canopy cover, can be incorporated in $\bm{h}_1$ as:
\begin{align}
\label{eqn:18}
\bm{h}_{1}=\bm{U}\bm{x}_{1+\triangle t} +\bm{U}_2 \bm{z_{slp}}+\bm{U}_3 \bm{z_{asp}}+\bm{U}_4 \bm{z_{veg}}+\bm{U}_5 \bm{z_{cnp}}, 
\end{align}
where $\bm{Z}=[\bm{z}_{slp},\bm{z}_{asp},\bm{z}_{veg},\bm{z}_{cnp}]$ denote covariates for slope, aspect, existing vegetation cover, and forecast canopy cover, while $\bm{U}_2,\bm{U}_3,\bm{U}_4,\bm{U}_5$ are pre-specified (random) weight matrices. When considering such a model, we did not observe significant improvement in forecasting with (\ref{eqn:18}) and one cannot perform inference on the importance of these effects directly, as is possible in the fully hierarchical Bayesian framework in \citet{yoo}. Alternatively, one can add an additional stage in (\ref{eqn:eqn9}) to concatenate covariates in hidden states as:
\begin{align}
 \bm{h}_{t-\triangle t}^*= [\bm{h}_{t-\triangle t}^{\top},\bm{z}^{\top}]^{\top},
\end{align}
where $\bm{z}^{\top}=[z_{slp}^{\top},z_{asp}^{\top},z_{veg}^{\top},z_{cnp}^{\top}]^{\top}$. Even though the readout matrix corresponding to $(19)$ can quantify the element-wise effect of covariates, this model did not show significant improvement in forecasting. Therefore, a subject of future research is to consider alternative approaches for introducing covariates into our hybrid model in a manner that can provide explainability regarding covariate importance \citep[e.g.,][]{wikle2022illustration}. 

Another direction in which to extend our model is to account for topological constraints explicitly. Although we demonstrate in Figure \ref{fig3} that our model effectively accounts for such constraints implicitly when there is data to inform the normal speed, in situations where there are data gaps in time, or multiple time step forecasts are of interest, then explicit constraints would be important to produce realistic forecasts (e.g., accounting for water bodies). 
Employing a re-initialization technique in the model could be another path. \cite{osher2} suggested occasional re-initialization to maintain a signed distance function. Nevertheless, we observe that re-initialization is not necessary for one-step ahead forecasting given data at each time step (see Figure \ref{fig3}, \ref{fig5}, and \ref{fig7}). Yet, for multiple-step ahead forecasting, re-initialization would likely be beneficial.

Lastly, it would be instructive to combine the level set method with other variants of ESNs. Our model is based on the basic ESN, but we could also consider quadratic or deep ESNs  \citep[e.g.,][]{mcdermott,McDermott2}.

\bibliographystyle{apalike}
\bibliography{references}

\end{document}